# Quantum manipulation of two-color stationary light: Quantum wavelength conversion

S. A. Moiseev[1,2] and B. S. Ham[1]

[1]*The Graduate School of Information and Communications, Inha University, Incheon 402-751 S. Korea*
[2]*Kazan Physical-Technical Institute of Russian Academy of Sciences, Kazan 420029 Russia*

We present a quantum manipulation of a traveling light pulse using double atomic coherence for two-color stationary light and quantum frequency conversion. The quantum frequency conversion rate of the traveling light achieved by the two-color stationary light phenomenon is near unity. We theoretically discuss the two-color stationary light for the frequency conversion process in terms of pulse area, energy transfer and propagation directions. The resulting process may apply the coherent interactions of a weak field to nonlinear quantum optics such as quantum nondemolition measurement.

**Introduction**

   Coherent interactions have been intensively studied for the last several decades in nonlinear optics to utilize weak light for enhanced light-matter interactions [1]. Because nonlinear optics should intrinsically rely on strong field interactions, using a weak field light has been a fundamental limitation in pump laser intensity [2]. Thus, coherent interactions have drawn much attention to the weak-field limited nonlinear optics. The weak field nonlinear interactions are mostly important to quantum information sciences that need repeated non-destructive measurements of physical observables and deterministic quantum switching for quantum processing [3]. So far, an optical cavity has been an essential tool to control the weak quantum light interacting with an optical medium [4]. However, applying the optical cavity to the traveling light has been a major drawback in operational bandwidth due to the trade-off between the trapping time and the transparency of the cavity.

   Electromagnetically induced transparency (EIT) [5] has been studied for the last fifteen years to alleviate the intensity threshold in nonlinear optics and has been successfully applied to quantum optics by utilizing nonabsorption resonance and dispersion control of an optical medium in a traveling light scheme [6-10]. Actually the dispersion control of EIT is of great benefit to us when using the traveling light for deterministic quantum coherent control to decelerate or even completely stop the light pulses. With a simple modification of the EIT scheme, one can obtain a giant phase shift [11], where the phase shift itself has been a major goal in nonlinear optics for switching applications as well as in quantum optics for generation of Schrodinger's cat states [12]. Here, it should be noted that the π-phase shift is an essential acquirement for the optical switching based on Mach-Zehnder interferometer and quantum superposition. The giant phase-shift in a weak field limit has been studied using stationary light phenomenon [13]. Recently, Lukin *et al*. demonstrated photon trapping in a hot atomic gas using identical counterpropagating control fields resulting in standing-wave grating [14]. The photon trapping as a stationary light gives a predominant property to nonlinear quantum optics particularly when an optical medium is spatially limited and highly absorptive. The stationary light phenomenon can replace the conventional cavity QED with an advantage of no-high-Q mirrors. Unfortunately the original technique of Ref. [14] must be used with cold atomic gases (Bose-Einstein [BEC]) or solids. Based on the stationary light scheme, on-demand quantum entanglement generation has also been proposed [15].

   The scheme of the light trapping in Ref. [14] is an extension of the well-known idea of the photonic bandgap [16] to a slow light regime, where the light trapping results from strong Bragg scattering of the two counter-propagating slow light waves on the control-laser induced spatial grating. The strong interaction between the counter-propagating weak fields, however, can be effectively realized in the coherent medium due to the generation of an atomic coherence without additional spatial grating. The physical property of the quantum coherence is more manifold, so it can be principally used to control the stationary lights. In this paper we propose a method of dynamical manipulation of the stationary light: 1) to trap two or more quantum fields, particularly two-color (TC) light, and 2) to use the TC light for effective quantum wavelength conversion. The proposed scheme is based on the double coherence resulting from strongly coupled slow light through EIT, where it works with no-Doppler broadened media such as BEC or condensed media. We discuss the wavelength conversion process through the TC stationary light for the relationship between pulse broadening mechanism and energy transfer.

Figure 1 shows an energy level diagram for the present TC stationary light. Unlike intensity grating based stationary light scheme [13,14], we apply dichromatic control fields $\Omega_+$ and $\Omega_-$. As shown in Fig. 1, the propagation direction of the two control fields is opposite from each other, and the frequencies of these fields are $\omega_{C+}$ and $\omega_{C-}$ with detunings $\Delta_+ = \omega_{C+} - \omega_{32}$ and $\Delta_- = \omega_{C-} - \omega_{32} \neq \Delta_+$, respectively, where the frequency difference is $|\omega_{C+} - \omega_{C-}| \gg \delta\omega_p$ ($\delta\omega_p$ is a spectral width of the slow light pulses). The input quantum light $E_+$ is assumed to propagate in the $\mathbf{k_+}$ direction. Here, the propagation direction $\mathbf{k_-}$ of the field $E_-$ is determined by the phase matching with Bragg condition: $\mathbf{k_-} = \mathbf{k_+} - \mathbf{\kappa_+} + \mathbf{\kappa_-}$. Unlike intensity-grating-based standing light [14], the present scheme utilizes quantum coherence $\rho_{12}$ created by the Raman fields of $E_+$ & $\Omega_+$ and $E_-$ & $\Omega_-$. We assume that initially all the atoms are in the ground states $|1\rangle$, and the probe pulse $E_+$ enters into the medium in the presence of the intensive control field $\Omega_+$ only, which is co-propagating in the same direction $\mathbf{\kappa_+}$ as the probe field direction $\mathbf{k_+}$. We note the frequency $\omega_+$ of the field $E_+$ is detuned from the atomic transition $|1\rangle - |3\rangle$ by $\Delta_+ = \omega_+ - \omega_{31}$. For strong control field $\Omega_+$ the probe quantum field $E_+$ can decelerate through the medium if the condition of EIT is satisfied [5-10]. We analyze the space-time evolution of the quantum light when an additional intensive control field $\Omega_-$ is switched on. It is especially interesting for us to study how the new weak quantum field $E_-$ is generated in different frequencies, where the spectral width of the initial probe light satisfies $\delta\omega_\pm \ll \Delta_+ - \Delta_-$ for practical frequency conversion processes.

We show that the generated quantum field $E_-$ interacts so intensively with the original probe field $E_+$, so the coupled field of the TC light pulse is formed in the optically dense medium. We also prove that the TC light pulse gets some spatial structure and unified group velocity depending on the Rabi frequencies of the control fields ($\Omega_+(t)$ and $\Omega_-(t)$) and parameters of the atomic medium. It is interesting that the manipulation with the Rabi frequencies of the control fields directly determines both the amplitude and group velocity of each spectral component of the TC-light. The most striking result is the possibility of quantum stopping of the TC light inside the medium and ultra-efficient wavelength conversion, where the quantum state of the new $E_-$ field should be a time reversed replica of the probe pulse $E_+$.

**Theory of TC stationary light**

For a theoretical analysis of the two-color stationary lights we introduce quantum field $\hat{E}_\sigma = \sqrt{\hbar\omega_\sigma/(2\varepsilon_o V)}\hat{A}_\sigma e^{-i\omega_\sigma(t-\sigma z/c)} + H.C.$, where $\sigma = +/-$ stands for forward/backward field, respectively, which is traveling in a $\pm z$ direction, and where $A_\sigma$ is a slowly varying field operator. The interaction of the quantum fields with atoms driven by the control laser fields $\Omega_\pm(t) = \Omega_{\pm,o}(t)e^{i\phi_\pm}$ ($\Omega_{\pm,o}(t)$ and $\phi_\pm$ are slowly varied Rabi frequencies and phases of the fields) is given by the following Hamiltonian in the interaction picture:

$$\hat{H} = -\hbar g \sum_{j=1}\{\hat{A}_+(t,z_j)e^{-i(\Delta_+ t-k_+ z_j)} + \hat{A}_-(t,z_j)e^{-i(\Delta_- t+k_- z_j)}\}\hat{P}_{31}^j - \hbar\sum_{j=1}\{\Omega_+ e^{-i(\Delta_+ t-K_+ z_j)} + \Omega_- e^{-i(\Delta_- t+K_- z_j)}\}\hat{P}_{32}^j + H.C.,$$
(1)

where $\hat{P}_{nm}^j = (\hat{P}_{mn}^j)^+$ are the atomic operators, $g_\sigma = \wp_\sigma\sqrt{\omega_\sigma/(2\varepsilon_o\hbar V)}$ is a coupling constant of photons with atoms, $\wp_\sigma$ is a dipole moment for each transition, V is volume of quantization [1], $\Delta_\pm = \omega_\pm - \omega_{31}$, $k_\pm = \omega_\pm^p/c$, $K_\pm = \omega_\pm^c/c$.

Using Hamiltonian (1) we derive the equations for the atomic operators $P_{mn}^j$ and weak probe fields $\hat{A}_\pm$ adding the decay constants $\gamma_3$ and $\gamma_2$ for the atomic transitions $|1\rangle-|3\rangle$ and $|1\rangle-|2\rangle$:

$$\tfrac{\partial}{\partial t}\hat{P}_{13}^{j} = -\gamma_3 \hat{P}_{13}^{j} + ig\{\hat{A}_+(t,z_j)e^{-i(\Delta_+ t - k_+ z_j)} + \hat{A}_-(t,z_j)e^{-i(\Delta_- t + k_- z_j)}\} + i\{\Omega_+ e^{-i(\Delta_+ t - K_+ z_j)} + \Omega_- e^{-i(\Delta_- t + K_- z_j)}\}\hat{P}_{12}^{j} + \hat{f}_3(t,z_j),$$
(2)

$$\tfrac{\partial}{\partial t}\hat{P}_{12}^{j} = -\gamma_2 \hat{P}_{12}^{j} + i\{\Omega_+^* e^{i(\Delta_+ t - K_+ z_j)} + \Omega_-^* e^{i(\Delta_- t + K_- z_j)}\}\hat{P}_{13}^{j} + \hat{f}_2(t,z_j),$$
(3)

where $\hat{f}_{2,3,4}(t,z_j)$ are the Langevin forces associated with the relaxation processes [1]. We ignore the influences of the forces in Eqs. (2) and (3) for the following analysis in the adiabatic limit [16]. Eqs. (2) and (3) have been obtained within the limit of the weak quantum fields where the population of the level 3 and atomic coherence $\hat{P}_{23}$ are ignored [1]. Assuming both $|\Delta_+ - \Delta_-|\delta t_o \gg 1$ ($\delta t_o$ is a temporal duration of the probe pulse $E_+$) and that spectral widths of the lights $E_+$ and $E_-$ are smaller than their frequency difference, we apply $\hat{P}_{13} = \hat{P}_+ e^{-i(\Delta_+ t - k_+ z)} + \hat{P}_- e^{-i(\Delta_- t + k_- z)}$ into Eq. (2) and Eq.(3) to find simpler system equations for the operators $\hat{P}_\pm$ and $\hat{P}_{12}$:

$$\tfrac{\partial}{\partial t}\hat{P}_+ = -\gamma_+ \hat{P}_+ + ig\hat{A}_+ + i\Omega_+ e^{-ik_o z}\hat{P}_{12},$$
(4)

$$\tfrac{\partial}{\partial t}\hat{P}_- = -\gamma_- \hat{P}_- + ig\hat{A}_- + i\Omega_- e^{ik_o z}\hat{P}_{12},$$
(5)

$$\tfrac{\partial}{\partial t}\hat{P}_{12} = -\gamma_2 \hat{P}_{12} + i\{\Omega_+^* e^{ik_o z}\hat{P}_+ + \Omega_-^* e^{-ik_o z}\hat{P}_-\}.$$
(6)

where $k_o = \omega_{21}/c$, $\gamma_\pm = \gamma_3 - i\Delta_\pm$, the j-th is dropped out for convenience.

For a long enough temporal duration of the probe field pulses $\hat{A}_\pm$ so that $|\gamma_\pm|\delta t_o \gg 1$ we have $\hat{P}_\pm \cong i\gamma_\pm^{-1}\{g\hat{A}_\pm + \Omega_\pm e^{\mp ik_o z}\hat{P}_{12}\}$.

For long lived atomic coherence,

$$\tfrac{\partial}{\partial t}\hat{P}_{12} = -\frac{\beta_s}{\gamma_+ \gamma_-}\hat{P}_{12} - (\hat{F}_+ + \hat{F}_-),$$
(7)

where $\beta_s(t) = (\gamma_2 \gamma_+ \gamma_- + \gamma_- \Omega_{+,o}^2(t) + \gamma_+ \Omega_{-,o}^2(t))$, $\hat{F}_\pm = (\gamma_\pm^{-1}\Omega_\pm^* e^{\pm ik_o z}g\hat{A}_\pm)$.

The Eq. (7) is added by the equations for the field operators $\hat{A}_\pm$:

$$(\tfrac{\partial}{c\partial t} + \tfrac{\partial}{\partial z})\hat{A}_+ = i(Ng/c)\hat{P}_+,$$
(8)

$$(\tfrac{\partial}{c\partial t} - \tfrac{\partial}{\partial z})\hat{A}_- = i(Ng/c)\hat{P}_-.$$
(9)

We study the slow light evolution in the switching processes of the control laser fields $\Omega_\pm(t)$ under the typical adiabatic condition for the slow light propagations $\delta t_o |\beta_s(t)/(\gamma_+ \gamma_-)| \gg 1$, where $\hat{P}_{12} \cong \{-1 + (\gamma_+ \gamma_- / \beta_s)\tfrac{\partial}{\partial t}\}\gamma_+ \gamma_- (\hat{F}_+ + \hat{F}_-)/\beta_s$. After substitution $\hat{P}_{12}$ into Eq. (8) and (9) and introducing the field operators $\hat{\Psi}_\pm = e^{\pm ik_o z}\sqrt{N}g\hat{A}_\pm/\Omega_\pm(t)$ (N is atomic density) we obtain the following coupled wave equations under the condition of a slow-light propagation $v_o(t) \cong c\Omega_+^2(t<t_o)/Ng^2 \ll c$ ($v_o$ is group velocity of the initial probe field $E_+$):

$$\{\tfrac{\partial}{\partial z} - ik_o\}\hat{\Psi}_+ = -\gamma_2 \gamma_3 \gamma_- (\xi/\beta_s)\hat{\Psi}_+ - \xi\tilde{\alpha}_-(\hat{\Psi}_+ - \hat{\Psi}_-) - \xi\frac{\gamma_-}{\beta_s}\frac{\partial}{\partial t}(\gamma_-\tilde{\alpha}_+\hat{\Psi}_+ + \gamma_+\tilde{\alpha}_-\hat{\Psi}_-),$$
(10)

$$(\tfrac{\partial}{\partial z} + ik_o)\hat{\Psi}_- = \gamma_2 \gamma_3 \gamma_+ (\xi/\beta_s)\hat{\Psi}_- - \xi\tilde{\alpha}_+(\hat{\Psi}_+ - \hat{\Psi}_-) + \xi\frac{\gamma_+}{\beta_s}\frac{\partial}{\partial t}(\gamma_-\tilde{\alpha}_+\hat{\Psi}_+ + \gamma_+\tilde{\alpha}_-\hat{\Psi}_-).$$
(11)

where $k_0 = \omega_{21}z/c$, $\xi = Ng^2/c\gamma_3$ is the absorption coefficient on the transition $|1\rangle$-$|3\rangle$, $\tilde{\alpha}_\pm(t) = \gamma_3 \Omega_{\pm,o}^2(t)/\beta_s(t)$ and $\delta t|\beta(t)/(\gamma_+\gamma_-)| \gg 1$. Ignoring the absorption of the slow light before the interaction with new field $E_-$ ($\gamma_2 t_o \ll 1$), we note the operators $\hat{\Psi}_\pm(t,z)$ describe the coupled field and dark polariton excitations in the medium (in the adiabatic limit) where the atomic coherence $\hat{P}_{12}$ between the

states $|1\rangle$ and $|2\rangle$ becomes also proportional to field operators $A_\pm$. In such a way we can introduce the initial quantum state of the field at the time $t=t_o$: $|(t_o)_{in}\rangle = |\rho(t_o)\rangle \otimes |0\rangle_-$ where $|0\rangle_\pm$ is a vacuum state of the fields $\hat{\Psi}_\pm(t,z)$ and $|\rho(t_o)\rangle$ corresponds to the state of the slowly propagating field $\hat{\Psi}_+(t_o,z)$. In a particular case of a single photon state $|\rho_{ph}(t_o)\rangle = \int dk \rho_{ph}(k)\hat{\psi}^+_{+;k}|0\rangle_+$ ($\hat{\psi}_{+,k}(t_o) \equiv \hat{\psi}_{+;k}$), where $\int dk |\rho_{ph}(k)|^2 = 1$, and $\hat{\psi}^+_{\sigma;k}$ is a boson create operator of the ω-th mode (of the field $\Psi_\sigma(t_o,z)$) with the following commutation roles $[\hat{\psi}_{\sigma';k'}, \hat{\psi}^+_{\sigma;k}] = \delta_{\sigma',\sigma}\delta(k'-k)$ (see also [16]) for two weak quantum fields $\hat{\Psi}_\pm(t_o,z)$.

The system equations (10)-(11) can be solved analytically using spatial Fourier transformation $\hat{\Psi}_\sigma(t,z) = \int_{-\infty}^{\infty} dk e^{ikz} \hat{\psi}_{\sigma,k}(t)$. Assuming $\Omega_-(t<t_o) = 0$ in Eqs. (10) and (11) we find the field Fourier components $\hat{\Psi}_\pm(t,z)$ via the initial field operators $\hat{\psi}_{+,k}$:

$$\hat{\psi}_{\sigma,k}(t) = \chi_\sigma(k) B(k,k_o;t,t_o) \hat{\psi}_{+,k}, \tag{12 a}$$

$$B(k,k_o;t,t_o) = \frac{\gamma_- \tilde{\alpha}_+(t_o)}{[\gamma_- \tilde{\alpha}_+(t) + \gamma_+ \tilde{\alpha}_-(t)\chi_-(k)]} \exp\{-i\int_{t_o}^{t} \omega_{k,k_o}(t')dt'\}, \tag{12 b}$$

where $\chi_-(k) = \{1 + i(\gamma_+/\gamma_3)\delta k_-\}/\{1 - i(\gamma_-/\gamma_3)\delta k_+\}$, $\chi_+(k) = 1$, $\delta k_{+,-} = (k \pm k_o)/\xi$, the frequency $\omega_k(t)$ determines the nonstationary dispersion relation of the fields:

$$\omega_{k,k_o}(t') = -i\gamma_2 \frac{1+i\mu_{+-}(k)}{J(t,k,k_o)} + \beta_s(t)\frac{[\tilde{\alpha}_+(t)\delta k_- - \tilde{\alpha}_-(t)\delta k_+ - i\delta k_+ \delta k_-]}{\gamma_3^2 J(t,k,k_o)}, \tag{13}$$

where for simplicity we have used the following nodimensional functions: $\mu_{+-}(k) = [(\gamma_+/\gamma_3)\delta k_- - (\gamma_-/\gamma_3)\delta k_+]$ and $J(t,k,k_o) = \{1 - (\gamma_2\gamma_+\gamma_-)/\beta_s(t) + i[(\gamma_+/\gamma_3)^2 \delta k_- \tilde{\alpha}_-(t) - (\gamma_-/\gamma_3)^2 \delta k_+ \tilde{\alpha}_+(t)]\}$.

If only one of the control laser fields is switched on ($\Omega_\pm(t) \neq 0, \Omega_\mp(t) = 0$), we find the following from Eq. (13):

$$\omega_{k,k_o}(t, \Omega_\pm(t) \neq 0, \Omega_\mp = 0) = \frac{(\gamma_2\gamma_\pm + \Omega^2_{\pm,o}(t))}{\Omega^2_{\pm,o}(t)}\{-i\gamma_2 \pm [\gamma_2\gamma_\pm + \Omega^2_{\pm,o}(t)]\frac{c(k \mp k_o)}{Ng^2}\}\Big|_{\gamma_2 t \ll 1}$$

$$\cong -i\gamma_2 \pm c\frac{\Omega^2_{\pm,o}(t)}{Ng^2}(k \mp k_o), \tag{14}$$

which corresponds to the well-known limit of the slow light with a weak absorption. So the modes of the field $\hat{\varphi}^+_{+;k}$ ($\gamma_2 t_o \ll 1$, $t \approx t_o$) are characterized by the dispersion relation $\omega = v_o(k-k_o)$.

We note an interesting general connection between the Fourier components of the coupled light $\hat{\psi}_{-,k}(t) = \chi_-(k)\hat{\psi}_{+,k}(t)$, where the factor $\chi_-(k)$ is independent of the Rabi frequencies $\Omega_\pm(t)$ and of the decay constant $\gamma_2$. Such relation point out to the universally similar dynamics of the interacting field $\hat{\psi}_{\pm,k}(t)$. Applying inversed Fourier transformation to Eq. (12 a) we write the solution in the following form:

$$\hat{\Psi}_\pm(z, t > t_0) = \int dz' G_\pm\{(z-z'), t; t_o; k_o\}\hat{\Psi}_+(t_o, z'), \tag{15}$$

where Green functions $G_\pm\{(z-z'), t, t_o; k_o\}$ are:

$$G_\pm\{(z-z'), t, t_o; k_o\} = (2\pi)^{-1}\int_{-\infty}^{\infty} dk e^{ik(z-z')}\chi_\pm(k)B(k,k_o;t,t_o). \tag{16}$$

Applying inversed Fourier transformation to Eq. (12) we find that the fields $\hat{\Psi}_-(t,z)$ and $\hat{\Psi}_+(t,z')$ are coupled to each other inseparably in the same spatial domain for the arbitrary control field's parameters via the following nonlocal relation:

$$\hat{\Psi}_-(t,z) = \int_{-\infty}^{\infty} dz' \, \chi_{-;+}(z-z',k_o)\hat{\Psi}_+(t,z'), \qquad (17)$$

$$\chi_{-;+}(z-z',k_o) = \frac{\gamma_+}{\gamma_-}\{2\eta(z'-z)[\xi\gamma_3 \frac{(\gamma_+ + \gamma_-)}{2\gamma_+\gamma_-} - ik_o]\exp[(\xi\frac{\gamma_3}{\gamma_-} - ik_o)(z-z')] - \delta(z-z')\}, \qquad (18)$$

where $\eta(x)=1$ for x>0 and, $\eta(x)=0$ for x<0. Eqs. (15) and (17) point out that the quantum correlations between the fields $\hat{\Psi}_-(t,z)$ and $\hat{\Psi}_+(t,z')$ are spreading within in the spatial size about $l_{cor} = \xi^{-1}\sqrt{(\Delta_-^2 + \gamma_3^2)/\gamma_3^2}$ for arbitrary quantum states due to the sharp spatial behavior of the function $\chi_{-,+}(z-z',k_o)$ having one maximum near to z=z' with a spatial width $l_{cor}$. Therefore the fields $\hat{\Psi}_-(t,z)$ and $\hat{\Psi}_+(t,z')$ can be copies of each other in the optically dense medium if the spatial size $l_o$ of the initial probe light pulse is longer than $l_{cor}$, under the condition of $\Delta_-$ (or $\Delta_+$) ~ $\gamma_3$.

Using the solutions (12), (15) and (17) and initial quantum state $|(t_o)_{in}\rangle$ we can principally analyze the quantum properties of the fields $\hat{\Psi}_-(t,z)$ and $\hat{\Psi}_+(t,z')$ and their correlations. First of all we are interesting in the classical limit for average field amplitudes of the fields:

$$\langle \hat{A}_\sigma(t,z) \rangle = \frac{\Omega_\sigma(t)}{\sqrt{Ng^2}}\langle \hat{\Psi}_\sigma(t,z) \rangle, \qquad (19\,a)$$

where

$$\langle \hat{\Psi}_\sigma(t,z) \rangle = \langle (t_o)_{in}|\hat{\Psi}_\sigma(t,z)|(t_o)_{in}\rangle = \int_{-\infty}^{\infty} dk e^{ikz}\langle \hat{\psi}_{\sigma,k}(t) \rangle. \qquad (19\,b)$$

The integrals $\int_{-\infty}^{\infty} dk e^{ikz}\langle \hat{\psi}_{\sigma,k}(t) \rangle$ (in fact $\int_{-\infty}^{\infty} dk' e^{-ik'z'} \int_{-\infty}^{\infty} dk e^{ikz}\langle \hat{\psi}_{\sigma',k'}^+(t')\hat{\psi}_{\sigma,k}(t) \rangle$) cannot be calculated exactly due to the complicacy of the dispersion relation, so the complete analysis of the stationary lights needs numerical calculations.

Figure 2 shows numerical simulations of Eq. (19a) for the TC stationary light. As seen from Fig.2 by manipulations of the control laser fields $\Omega_\pm$ can work in trapping the quantum field E- as a timely-reversed quantum replica of the E+ propagating along the k- direction at different frequency: Quantum wavelength conversion. A more detailed explanation will be given in the following section.

**Manipulation with stationary light**

In this section, we analyze general properties of the coupled lights for some practical cases. First of all we obtain general information about the propagation of TC stationary light from the dispersion relation (13). An interesting situation in the quantum evolution of the lights takes place in the optically dense medium, where $|\delta k_{+,-}| = |(k \pm k_o)/\xi| << 1$. For weak relaxation $\gamma_2$ and small enough spectral detunings $|\gamma_\pm/\gamma_3| \sim 1$ we have $J(t,k,k_o) \approx 1$, $\mu_{+-}(k) << 1$ and $\beta_s(t) \cong (\gamma_-\Omega_{+,o}^2(t) + \gamma_+\Omega_{-,o}^2(t))$. Taking into account these simplifications we obtain the following relation from Eq. (13):

$$\omega_{k,k_o}(t) \approx -i\gamma_2 + \beta_s(t)\frac{[\tilde{\alpha}_+(t)\delta k_- - \tilde{\alpha}_-(t)\delta k_+]}{\gamma_3^2} = -i\gamma_2 + c\frac{[\Omega_{+,o}^2(t) - \Omega_{-,o}^2(t)]k}{Ng^2} - c\frac{[\Omega_{+,o}^2(t) + \Omega_{-,o}^2(t)]k_o}{Ng^2}. \qquad (20)$$

This formula describes an interesting property in the evolution: the two interacted fields $\hat{A}_+(t,z)$ and $\hat{A}_-(t,z)$ propagate together with one group velocity v(t):

$$v(t) = \frac{\partial}{\partial z}\omega_{k,k_o}(t) = c[\Omega_{+,o}^2(t) - \Omega_{-,o}^2(t)]/Ng^2. \qquad (21)$$

For chosen optical parameters we see the group velocity v(t) is independent of the frequency detunings $\Delta_\pm$ between the slow light frequencies $\omega_\pm$ and atomic frequency $\omega_{31}$ (see Fig.1). It is interesting to note that, due to this independence, the group velocity of the TC light coincides with the group velocity of coupled single frequency light in Ref. [14]. In accordance with Eq. (21) the united velocity of coupled TC light $v(t>t_o)$ can be easily controlled by manipulation with the control laser field amplitudes $\Omega_\sigma(t)$ ($\Omega_\sigma \neq 0$ for $t_o < t < t_1$). The united velocity $v(t)$ can be considerably decreased in comparison with the initial velocity $v_o$ and even the TC light can be completely stopped (v(t)=0) if $\Omega_+^2(t) = \Omega_-^2(t)$ (where $\widetilde{\alpha}_+(t) = \widetilde{\alpha}_-(t)$).

Using Eq. (12) we can analyze in detail the spatial and temporal properties of the TC stationary light evolution for arbitrary parameters of the initial probe pulse in the coherent state $|\rho_c(t_o)\rangle = \exp\{-(1/2)|A_{+,in}|^2 + A_{+,in}\hat{\psi}_+^+\}|0\rangle_+$ (where $\hat{\psi}_+^+ = \int_{-\infty}^{\infty} dk A(k)\hat{\psi}_{+;k}^+$, $\int_{-\infty}^{\infty} dk |A(k)|^2 = 1$) : Below we find analytically the most important information about the TC stationary light using the manipulation of the driving laser amplitudes $\Omega_+(t)$ and $\Omega_-(t)$ For simplicity we ignore weak decay rate $\gamma_2$ between the states |1> and |2> and consider small level splitting $\omega_{21} \ll 2c/(\xi l_o^2)$ in the case of ignorance of the phase mismatch between the fields $E_+$ and $E_-$. Taking into account such limitations in Eq. (13) we obtain the following dispersion relation:

$$\omega_{k,k_o}(t)\big|_{k_o \ll l^{-1},\xi} \cong \beta_s(t)\left\{\frac{\{\widetilde{\alpha}_+(t) - \widetilde{\alpha}_-(t)\}k - ik^2/\xi}{\gamma^2\{\xi + i[\widetilde{\alpha}_-(t)(\gamma_+/\gamma)^2 - \widetilde{\alpha}_+(t)(\gamma_-/\gamma)^2]k\}}\right\}. \quad (22)$$

We assume a Gaussian spectral shape in the initial state $|\rho_c(t_o)\rangle$:

$$A(k) = (l_o/\sqrt{\pi})^{1/2} \exp\{-\tfrac{1}{2}(kl_o)^2 + ik(z_o - v_o t_o) + i(\vartheta_+ - \phi_+)\}, \quad (23)$$

(where $l_o = v_o \delta t_o$, $z_o$ is coordinate of the probe field $E_+$ at t=$t_o$, $\vartheta_+$ is a constant phase). Using Eq. (19) we find the field amplitude of the probe $E_+$ before the second control field $\Omega_-(t)$ switching on:

$$\langle \hat{A}_+(t_o,z)\rangle = \tfrac{\Omega_\sigma(t)}{\sqrt{Ng^2}}\langle \rho_c(t_o)|\hat{\Psi}_+(t,z)|\rho_c(t_o)\rangle = A_{+;o}\exp\{+i\vartheta_+ - \tfrac{1}{2}(z + z_o - v_o t)^2/l_o^2\}, \quad (24\,a)$$

$$\langle A_-(t_o,z)\rangle = 0, \quad (24\,b)$$

where $A_{+,o} = \sqrt{\frac{2\sqrt{\pi}}{c\delta t_o}} A_{+,in}$ is the amplitude of the envelope which is independent of the group velocity of a slow light, $l_o$ is the spatial size of the pulse. Ignoring small terms proportional to $ck^3/\xi^2$ in Eq. (22) we get the solutions for $\langle \hat{A}_\sigma(t,z)\rangle$ for arbitrarily varying control laser fields $\Omega_\sigma(t)$ after integration of Eq.(12) which is similar to Eq. (24) and taking into account the initial state $|\rho_c(t_o)\rangle$ (see Eq. (23)) and using Eq. (19) we find the following solution for the electromagnetic field amplitudes:

$$A_\sigma(t,z) = \langle \hat{A}_\sigma(t,z)\rangle = A_{+;o}\frac{\Omega_\sigma(t)}{\Omega_+(t_o)}\frac{l_o}{l(t)}\exp\{+i\vartheta_+ - \tfrac{1}{2}[z + z_o - v_o t_o - z_\sigma(t)]^2/l^2(t)\}, \quad (25\,a)$$

where

$$z_\sigma(t) = \sigma\gamma_\sigma/(\gamma\xi) + \int_{t_o}^{t} v(t')dt', \; v(t') = \beta(t')[\widetilde{\alpha}_+(t') - \widetilde{\alpha}_-(t')]/(\gamma^{-2}\xi), \quad (25\,b)$$

and

$$l(t) = \sqrt{[l_o^2 + 2\chi(t)/\xi^2]}, \; \chi(t) = \gamma^{-2}\int_{t_o}^{t} dt'\beta(t')M(t'),$$

$$M(t') = \{1 - [\widetilde{\alpha}_+(t') - \widetilde{\alpha}_-(t')][\widetilde{\alpha}_+(t')(\gamma_-/\gamma)^2 - \widetilde{\alpha}_-(t')(\gamma_+/\gamma)^2]\}. \quad (25\,c)$$

In accordance with the exact result of Eq. (17), we see that Eq. (25) demonstrates that the fields $A_+(t,z)$ and $A_-(t,z)$ move together. However, the fields are separated from each other by a constant distance $D = z_+(t) - z_-(t) = (\gamma_+ + \gamma_-)/(\gamma\xi) \ll l_o$, which is independent of the group velocity v(t). It is interesting that the distance D can be minimized for symmetric detunings of the control laser fields $\Delta_+ = -\Delta_-$, where $D = D_{min} = 2/\xi$. From Eq. (25c) we get the spatial size $l(t)$ of the TC stationary pulses for v(t)=0 and

$\Omega_+^2 = \Omega_-^2$:

$$|l(t)| = \sqrt[4]{\{[l_o^2 + 4v_o(t-t_o)/\xi]^2 + 4[v_o(t-t_o)(\Delta_+ + \Delta_-)/(\xi\gamma_3)]^2\}} \cdot \qquad (26)$$

This spreading of the pulse envelope $d|l(t)|/dt$ is a minimum for the symmetric spectral detuning $\Delta_+ = -\Delta_-$ again where the size $l(t) = \sqrt{l_o^2 + 4v_o(t-t_o)/\xi}$ and the stationary light evolves with minimal spatial distortion for arbitrary spatial envelopes. Thus the symmetric detuning condition $\Delta_+ = -\Delta_-$ is the most convenient factor for stable manipulations with TC light. We note that imaginary parts of $z_\sigma(t)$ determines a stationary phase modulation of the field envelopes, which is proportional to $-\sigma\Delta_\sigma/(\gamma_3 \xi l(t)) < 1$ for v(t)=0 and is negligible in the optically dense medium. The stationary TC light evolution in space and time is demonstrated in the Fig. 3 for the symmetric condition ($\Delta_+ = -\Delta_-$) and stopping condition $\Omega_+ = \Omega_-$. As seen in Fig.3, the pulse spreading results from the interaction of the TC light with the medium. Fig.3 concludes that the pulse area should be conserved if $\gamma_2$=0, even though there is pulse spreading.

In accordance with Eq. (25 a) the coupled quantum fields $A_\sigma$ are generated only when both control fields, $\Omega_+$ and $\Omega_-$, are turned on. The envelopes, A$_+$(t,z) and A$_-$(t,z), approximately overlap with each other. For the stationary light condition, $\Omega_+ = \Omega_-$ (t$_0$<t<t$_1$), the traveling light A$_+$ comes to a stand-still until $\Omega_-$ is turned off: Figures 2(b) and 2 (f) demonstrate that the field A$_-$ is generated under the same condition.

Here we should point out that the stationary field amplitudes $A_\sigma$ (t,z) are proportional to the control field $\Omega_+(t) = \Omega_-(t) = \Omega(t)$, so the quantum field amplitudes can be considerably amplified by the ratio $\Omega(t)/\Omega_+(t_o) > 1$, where the magnitude of $\Omega_+$ (t$_0$) is different from $\Omega_+$ (t$_0$<t<t$_1$). The amplification is simply explained with the atomic coherence built up during the slow light procedure. When a light pulse experiences a slow group velocity, the energy of the light pulse should be reduced roughly by the ratio of v$_g$/c. Thus, by the energy conservation law, the amount of lost energy must be converted into the atomic coherence $\rho_{21}$ (will be discussed elsewhere). The amplification is from the retrieval of the atomic coherence, but cannot be larger than the initial energy of the probe field E$_+$. Such results can be effectively used for enhancement of the nonlinear interactions of single photon fields interacting with collective ensemble [13,17].

By turning off the control pulse $\Omega_-(t) \to 0$ at t=10, the field A$_-$ disappears completely (see Fig. 2(b)), and the field A$_+$ starts to propagate in its original direction (see Fig. 2(a)). In the opposite situation, where $\Omega_+(t)$ is switched off, the field A$_+$ disappears but the quantum field $A_-$ now propagates backward with new carrier frequency ω$_-$ (see Fig. 2(c) and (d)).

Using Eq. (12) we get the following exact solutions for the field amplitudes $\langle A_\pm(t>t_1,z)\rangle$ after the control field switching operations:

$$\langle \hat{A}_\pm(t>t_1,z)\rangle = \tfrac{\Omega_\sigma(t)}{\sqrt{N}g^2}\int dz' G_{\Omega_+(t>t_1)=0}\{(z-z'),t;t_o,k_o\}e^{\mp ik_o z'}\langle\hat{\Psi}_+(t_o,z')\rangle, \qquad (27)$$

where Green functions are:

$$G_{\Omega_+(t>t_1)=0}\{(z-z'),t>t_1;t_o,k_o\} = (2\pi)^{-1}\int dk \exp\{k[(z-z')]-i\int_{t_o}^{t}\omega_{k,k_o}(t')dt'\}_{\Omega_+(t>t_2)=0}, \qquad (28)$$

Using Eq. (14) we find the relation between the field envelope areas under the slow light field A$_+$(t,z) as well as for A$_-$(t,z) after $\theta_\pm(t>t_1)$ and before $\theta_+(t_o)$ the switching process $\theta_\pm(t) = \int dt\langle A_\pm(t,z)\rangle \cong v_\pm^{-1}\int_0^L dz\langle A_\pm(t,z)\rangle$:

$$\theta_\pm(t>t_1) = e^{-i\gamma_2(t-t_o)}\theta_+(t_o)\exp\{-i\varepsilon_\pm(t_1,t_o)\}, \qquad (29)$$

where $\varepsilon_{\pm k_o}(t_1, t_o) = \int_{t_o}^{t_1} \omega_{k=\pm k_o, k_o}(t')dt'$. Assuming stationary TC light for $t_o < t < t_1$ ($\Omega_{+,o}^2 = \Omega_{-,o}^2 \equiv \Omega^2$) we get

$$-i\varepsilon_{\pm k_o}(t_1, t_o) = (t_1 - t_o)\left\{-\gamma_2 + 2\frac{\Omega^2}{\gamma_3}\frac{k_o}{\xi}\left(i - 2\frac{k_o}{\xi}\frac{(\gamma_{\mp}^*)^2(\gamma_+ + \gamma_-)}{\gamma_3|\gamma_+ + \gamma_-|^2}\right)\left|1 - 2i\frac{k_o}{\xi}\frac{\gamma_{\mp}^2}{\gamma_3(\gamma_+ + \gamma_-)}\right|^{-2}\right\}. \quad (30)$$

By assuming very small $k_o$ value we find $\left|\theta_{\pm}(t > t_2)\right|_{\substack{\Omega_{\pm}(t > t_2) = 0 \\ k_o << \delta k, \xi}} \cong \left|\theta_{\pm}(t_o)\right|\exp\{-\gamma_2(t - t_o)\}$: The numerical demonstration is shown in Figs. 3 (a), (b) and (e) for conservation of the pulse areas for $\gamma_2 = 0$. This property points out the possible manipulations with the weak fields $A_{+,-}$ on a large temporal scale. Here, it should be noted that the maximum trapping time of the TC stationary light is determined by the coherence decay time $\gamma_2^{-1}$ between the two lowest energy states |1> and |2> (see Eq. (13) taking into account $\mu_{+-}(k) << 1$ and $J(t, k, k_o) \cong 1$ in the optically dense medium). For fast switching processes we get $\exp\{-i\int_{t_o}^{t_1}\omega_{k,k_o}(t')dt'\} \cong 1$, which leads to the following result (see Eqs. (12) and (12 b)):

$$\hat{\psi}_{\pm,k}(t_1)\bigg|_{\Omega_{\pm} \neq 0, \Omega_{\mp} = 0} \cong \hat{\psi}_{+,k}. \quad (31)$$

Taking into account Eq. (25) we conclude that Eq. (31) means nearly perfect quantum wavelength conversion for $\Omega_-(t > t_1) \neq 0$, where the new quantum field $E_-$ gets a quantum state of the initial state $|\rho(t_o)\rangle$. We note that the field $E_-$ is generated via the four-wave mixing processes, so the temporal profile (and spectral profile) of field $E_-$ is the reversed replica of the $E_+$. Fig. 3(g) shows the nearly perfect energy conversion for the wavelength conversion. Figs. 3(b)~(h) show numerical simulation for the pulse area and energy with non-zero decay rate $\gamma_2$. Non-zero $\gamma_2$ contributes to the energy loss even in the non-stationary region.

An interesting unexpected situation with the quantum state of the two fields $\hat{\Psi}_-(t, z)$ and $\hat{\Psi}_+(t, z')$ takes place in the transient stage when both control laser fields $\Omega_+$ and $\Omega_-$ coexist. For the particular case of a single photon initial state $|\rho_{ph}(t_o)\rangle$, we get quantum superposition of the two states corresponding to TC stationary light. In this case we have the following nonclassical properties of the field operators

$$\langle\hat{A}_{\sigma'}^+(t', z')\hat{A}_{\sigma}(t, z)\rangle = \langle\rho_{ph}(t_o)|\hat{A}_{\sigma'}^+(t', z')|0\rangle\langle 0|\hat{A}_{\sigma}(t, z)|\rho_{ph}(t_o)\rangle = A_{\sigma'}^*(t', z')A_{\sigma}(t, z) \neq 0, \quad (32\text{ a})$$

$$\langle\hat{A}_{\sigma'}(t', z')\hat{A}_{\sigma}(t, z)\rangle = 0, \quad (32\text{ b})$$

which show a quantum interference between the fields with different carrier frequencies and points out the possibility of detecting a photon only in one of the fields. Using Eqs. (12) and (15) (see also (25 a) and Figs. 3 (e) ~ (h)) we should note that the quantum wavelength conversion efficiency of the single photon field is limited ONLY by spatial broadening of the stationary light according to the following relation ($\gamma_2(t_1 - t_o) << 1$):

$$P(t > t_1; \omega_+ \to \omega_-)\bigg|_{\substack{\Omega_+ = 0 \\ \Omega_- \neq 0}} = \int_0^L dz\langle\hat{\Psi}_-(t_1, z)\hat{\Psi}_-(t_1, z)\rangle = l_o / l(t_1) \cong \left(1 + 4l_o^{-2}\frac{v_o}{\xi}(t_1 - t_o)\right)^{-1/2}. \quad (33)$$

If the trapping time satisfies the condition $t_1 - t_o << l_o^2\xi/(4v_o)$, the initial quantum field transforms to the new quantum field with probability close to unity: $P(\omega_+ \to \omega_-) \cong 1 - 4l_o^{-2}\frac{v_+}{\xi}(t_1 - t_o) \approx 1$. We note that Eq. (33) also determines the transformation of the average photon number to the new field (see Figs. 3 (g) and (h)).

**Conclusion**

In conclusion, we have presented the stationary two-color light and quantum wavelength conversion using double quantum coherence in a virtual double-Λ system. In this scheme, a traveling quantum field can be classically manipulated by simply adjusting the control fields' parameters for (1) deterministic TC stationary light, (2) selection of propagation direction – either backward or forward, and

(3) dynamic quantum wavelength conversion. The present quantum manipulation of two-color stationary may be applied to the quantum node where quantum storage and quantum switching are required.

For potential applications, semiconductor quantum wells or quantum dots may be considered as candidates because these structures have strong dipole moments and sharp absorption linewidth with a large absorption coefficient. The ultrashort decay time of sub-nanoseconds has become an essential feature of high-speed photonic devices. In general semiconductor quantum wells or quantum dots also have fast decoherence (dephasing) times of picoseconds, which is a critical limitation of quantum applications. This critical problem of fast decoherence time, however, can be solved by adjusting barrier thickness in coupled quantum wells for intersubband transitions.

**Acknowledgement**

This works was supported by Korea Research Foundation Grant KRF-2003-070-C00024. The authors also give thanks to Y. Chen for assistance with the computer simulations.

**References**
[1] E. Arimondo, *Progress in Optics XXXV*, Ch. 5 Ed. E. Wolf (Elsevier, New York, USA 1996); M. O. Scully and S. M. Zubairy, *Quantum Optics* (Cambridge Univ. Press, Cambridge, UK, 1997).
[2] Y. R. Shen, *The Principles of Nonlinear Optics* (John Wiley & Sons, New York, USA 1984).
[3] M. A. Nielsen and I. L. Chuang, *Quantum Computation and Quantum Information* (Cambridge Univ. Press, Cambridge, UK, 2000).
[4] J. Mckeever et al., Science **303**, 1992 (2004), and references therein.
[5] M. Fleischhauer, A. Imamoglu, and J. P. Marangos, Rev. Mod. Phys. **77**, 633 (2005), and references therein.
[6] L. V. Hau, S. E. Harris, Z. Dutton, and C. H. Behroozi, Nature **397**, 594 (1999).
[7] C. Liu, Z. Dutton, C. H. Behroozi, adn L. V. Hau, Nature **409**, 490 (2001).
[8] D. F. Philips et al., Phys. Rev. Lett. **86**, 783 (2001).
[9] O. Kocharovskaya, Y. Rostovtsev, and M. O. Scully, Phys. Rev. Lett. **86**, 628 (2001).
[10] A. V. Turukhin, V. S. Sudarshanam, M. S. Shahriar, J. A. Musser, B. S. Ham, and P. R. Hemmer, Phys. Rev. Lett. **88**, 023602 (2002).
[11] H. Schmidt and A. Imamoglu, Opt. Lett. **21**, 1936 (1996).
[12] D. Petrosyan and G. Kurizki, Phys. Rev. A **65**, 33 833 (2002); M. Paternostro, M. S. Kim, B. S. Ham, Phys. Rev. A **67**, 023811 (2003).
[13] A. Andre, M. Bajcsy, A. S. Zibrov, and M. D. Lukin, Phys. Rev. Lett. **94**, 063902 (2005).
[14] M. Bajcsy, A. S. Zibrov, and .M. D. Lukin, Nature (London) **426**, 638 (2003).
[15] S. A. Moiseev and B. S. Ham, Phys. Rev. A **72**, 053802 (2005).
[16] M. Fleischhauer, and M. D. Lukin, Phys. Rev. Lett. **84**, 5094 (2000).
[17] S. A. Moiseev and B. S. Ham, 'Manipulation of Two-Color Stationary Light Using Coherence Moving Gratings,' quant-ph/ 0512042 (2005).

Captions of figures

FIG. 1. An energy level design of for the two-color stationary light. The inset is a scheme of the wavelength conversion.

FIG. 2. Numerical simulations of the two-color stationary light. $A_+$ and $A_-$ are the amplitude of the two-color stationary light $E_+$ and $E_-$, respectively. (a) and (b) The control laser $\Omega_+$ is always on, while the $\Omega_-$ is on for $t_0$ (5) < t < $t_1$ (10). (c) and (d): The control laser $\Omega_+$ is on for 0 < t < $t_0$ (5), while the $\Omega_-$ is on $t_0$ (5) < t < 15. For. (e) and (f): Amplitudes of $A_+$ and $A_-$ for (a) and (b), respectively. The medium's optical constant is $\xi_+L=100$ ($\xi_+l_o=10$), where $l_o$ is the initial spatial size of the propagating quantum pulse and L is the medium's total length. The initial group velocity $v_g$ of the signal $E_+$ is 1. The temporal shape of the $A_+$ (when it enters into the medium) is Gaussian with a temporal duration of $T=1$. The spatial size is $l=v_gT$. $\gamma=100$; $\Omega_+=\Omega_-=100$; $\gamma_2=0$; $\Delta_+=\Delta_-$; $g_+/g_- = \Omega_+/\Omega_- = 1$.

FIG. 3. Numerical simulations of space-time evolution of the quantum field $E_+$ ($A_+$) for the case of Fig. 2 (a) for pulse area ((a), (b), (d), and (e)) and for energy ((c) and (f)). (a)-(c) $\gamma_2=0$ and (d)-(f) $\gamma_2=0.01$.

FIG. 1

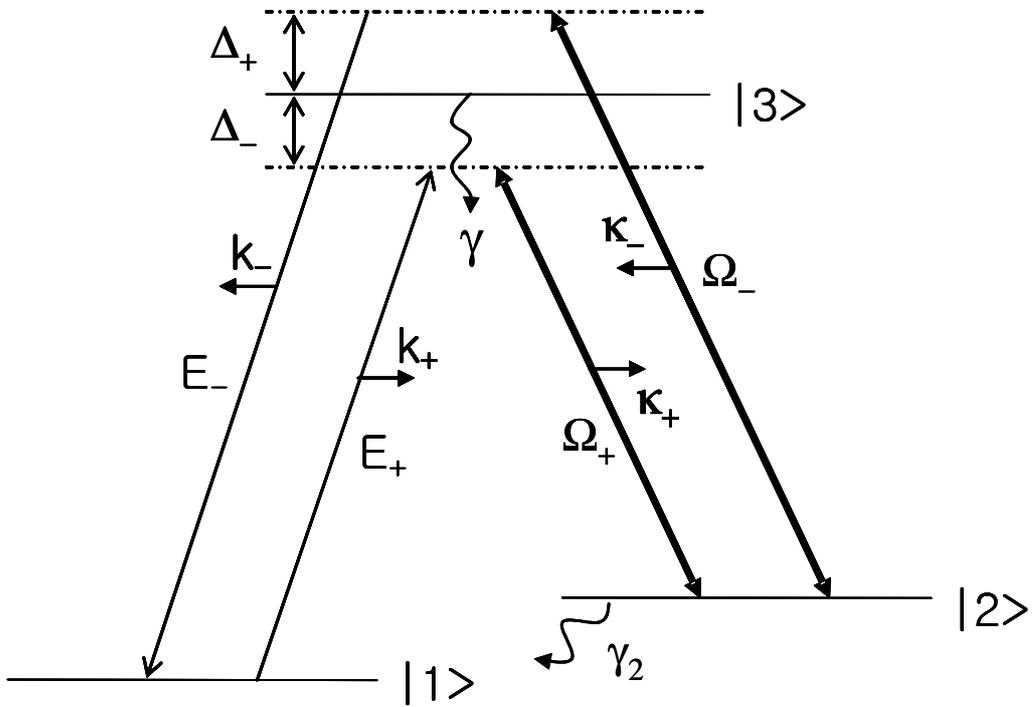

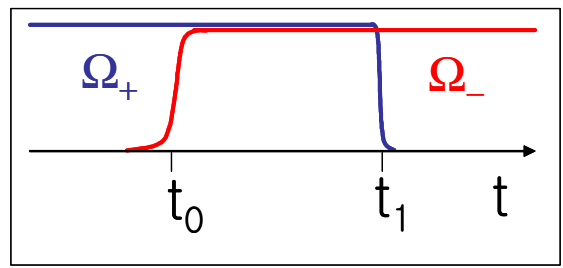

FIG. 2

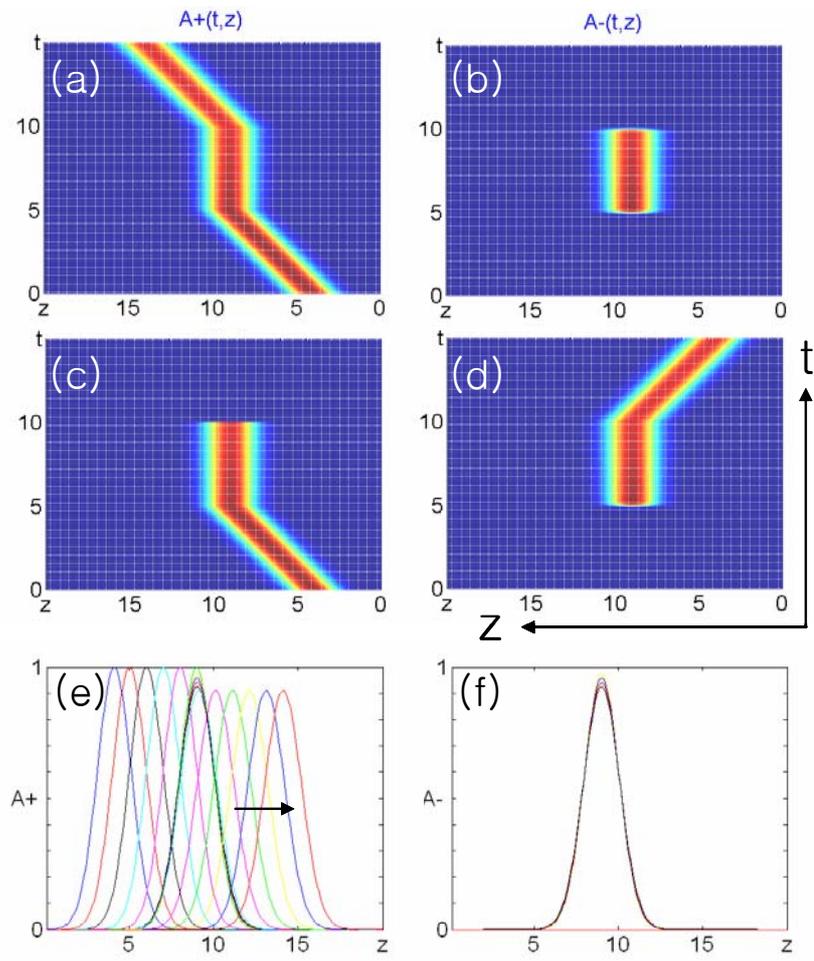

Fig. 3

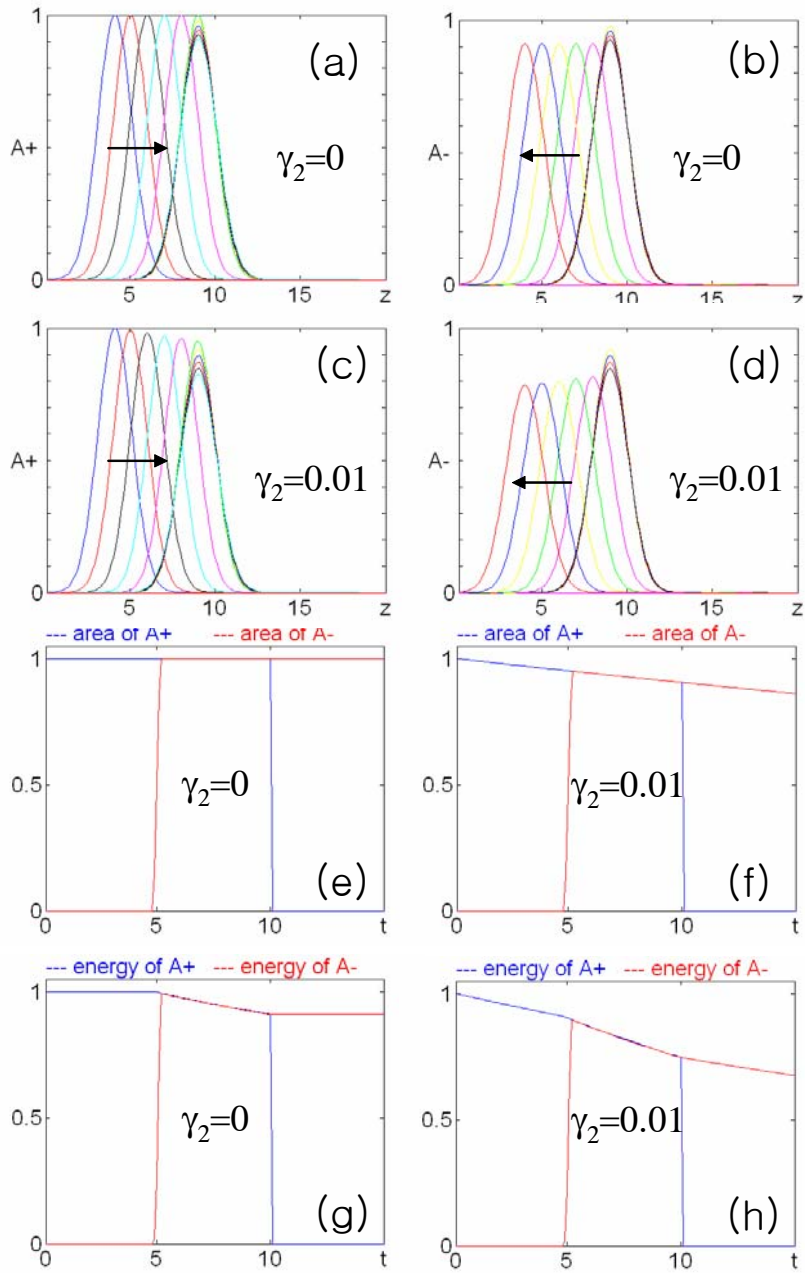